\documentclass[debug,overfull]{modepjb}

\usepackage{graphicx}
\usepackage{amsmath}
\usepackage{amsfonts}
\usepackage{amssymb}

\newcommand{\rr}{{\bf r}}

\newcommand{\ene}[1]{\epsilon_{#1}^{(e)}}
\newcommand{\enh}[1]{\epsilon_{#1}^{(h)}}
\newcommand{\enj}[2]{\epsilon_{#2}^{(#1)}}

\newcommand{\absv}[1]{\,| #1 |\,}

\newcommand{\veeMA}[4]{ V^{(e e)}_{#1 #2 #3 #4} }

\newcommand{\vhh}[4]{ V^{(h h)}_{#1 #2 #3 #4} }
\newcommand{\veh}[4]{ V^{(e h)}_{#1 #2 #3 #4} }
\newcommand{\vij}[6]{ V^{(#1 #2)}_{#3 #4 #5 #6 } }

\title{The carrier ''antibinding'' in quantum dots: a charge separation effect}

\author{Monique Combescot\inst{1} \and Marc-Andr\'{e} Dupertuis\inst{2}}
\institute{
\inst{1}Institut des Nanosciences de Paris, Universit\'{e} Pierre et Marie Curie and
Universit\'{e} Denis Diderot, CNRS, Campus Boucicaut,
140 rue de Lourmel, F-750015 Paris, France\\
\inst{2}Institute of Quantum Electronics and Photonics,
Faculty of Basic Sciences, Swiss Federal Institute of Technology-EPFL,
PHB-Ecublens, CH-1015 Lausanne, Switzerland}

\begin{document}
\maketitle

\begin{abstract}
We show that the carrier ''antibinding'' observed recently in 
semiconductor quantum dots, i.e., the fact that the ground state 
energy of two electron-hole pairs goes above twice the ground-state 
energy of one pair, can entirely be assigned to a charge separation 
effect, whatever its origin. In the absence of external electric field, this charge 
separation comes from different ''spreading-out'' of the electron and 
hole wavefunctions linked to the finite height of the barriers. When 
the dot size shrinks, the two-pair energy always stays below when the 
barriers are infinite. On the opposite, because barriers are less 
efficient for small dots, the energy of two-pairs in a dot with 
finite barriers, ends by behaving like the one in bulk, i.e., by 
going above twice the one-pair energy when the pairs get too close. 
For a full understanding of this ''antibinding'' effect, we have also 
reconsidered the case of one pair plus one carrier. We find that, 
while the carriers just have to spread out of the dot differently for 
the ''antibinding'' of two-pairs to appear, this ''antibinding'' for 
one pair plus one carrier only appears if this carrier is the one 
which spreads out the less. In addition a remarkable sum rule exists 
between the "binding energies" of two pairs and of one pair plus one carrier.
\end{abstract}





A very large amount of works are still devoted to the study of 
semiconductor quantum dots because of their possible applications in 
nanotechnology. The fundamental aspects of these quantum dots are 
however now essentially understood~\cite{Yoffe02}: when a few 
carriers of mass $m$ are confined in a box of characteristic size 
$R$, their kinetic energy is of the order of $\hbar^2 / m R^2$, while 
their Coulomb energy is order of $e^2 / R$; so that, if the box size 
is small compared to $\hbar^2 / m e^2$ (the so-called 
''strong-confinement regime''), Coulomb effects play a minor role --- 
even if the absolute value of the Coulomb energy in a dot is larger 
than the usual one in bulk, for the carriers are closer. This is why 
the physics of quantum dots is essentially a one-body physics, driven 
by confinement: besides small energy shifts and level splittings, 
many-body effects in a dot are not expected to be of great interest 
in these confined systems.

Recently, however, a rather surprising ''antibinding'' effect has 
been observed in these dots: if one measures the lowest energy of two 
electron-hole pairs in the strong confinement regime, one finds that 
it goes from below to above twice the ground state energy of 
one-pair, when the dot size decreases (see 
refs~\cite{Rodt03,Rodt05,Leni04,Kako04} and references therein). Let 
us stress that this is not really an ''antibinding'' effect because 
the carriers always stay bound to the dot due to the strong 
confinement. A two-pair energy above twice one-pair is however 
surprising at first because we are used to biexciton always having an 
energy below twice the exciton energy. This actually comes from the 
fact that, in extended systems, the excitons can move freely; so 
that, to decrease their energy, they adjust their distance at an 
optimum value $D^*$ which results from the competition between the 
kinetic energy they lose and the Coulomb energy they gain when they 
get closer.

The same argument may actually lead to think that the observed 
''antibinding'' is in fact just normal ! Indeed, if the particles get 
closer than $D^*$, which is what happens in small dots, the energy of 
two-pairs in bulk should start to rise because of the kinetic 
contribution. It should thus end by getting above twice the energy of 
one exciton. Consequently, it may appear as reasonable to find a 
two-pair ''antibinding'' when the dot size decreases, the pairs 
ending by being too close.

This way of thinking is actually incorrect: in a dot, the carriers 
are forced to stay together, at a given distance, by confinement. 
They have no choice ! The kinetic energy necessary to stay so close, 
is actually paid once we put the carriers in the box. When comparing 
the energy of two pairs to twice the energy of one pair, we are thus 
left with the Coulomb parts only. As the dipolar attraction between 
electron-hole pairs makes their Coulomb contributions to the energy 
always negative, this should lead to a two-pair energy always below 
twice the energy of one pair, in contradiction with the experimental 
data.

The purpose of this paper is to show that the energy of two pairs 
going above twice the energy of one pair can entirely be assigned to 
charge separation, whatever its origin. It must be pointed out that 
such a charge separation exists even in the 
absence of an external electric field. It results from a 
''spreading-out'' effect which increases when the dots shrink. The 
pairs, forced to stay closer than their optimum distance $D^*$, would 
love to get out of the box, in order to behave like free pairs in a 
bulk sample. This is of course impossible if the barrier height is 
really infinite: for such a barrier, the two-pair energy always stays 
below twice one-pair. However, for finite barriers, the carriers can 
partly escape from the dot and experience a subtle interplay between 
Coulomb interaction and confinement, i.e., interaction with the 
continuum linked to the environment of the dot~\cite{Michelini04}; 
Consequently, the price  in kinetic energy needed to put a carrier 
inside the dot is not really constant but depends on the dot size, 
through a {\em barrier-dependent} term.

In confined systems, what is really important is not so much the 
absolute value of the barrier height, but its relative value compared 
to the characteristic energy of the dot, namely $\hbar^2 / m R^2$. 
This led us to introduce~\cite{Leyronas01} the dimensionless 
parameter $\nu_i$  which characterizes a barrier of height $V_i$ for 
a carrier of mass $m_i$ trapped in a spherical dot of radius $R$. 
This parameter is defined as
\begin{equation}\label{eq:Vi}
V_i = \frac{\nu_i^2\hbar^2}{2 m_i R^2}
\end{equation}
While  $\nu_i$ is always infinite when $V_i$ is infinite, it goes to 
zero for finite $V_i$ when the dot shrinks : A dot size reduction 
makes a given barrier less and less efficient to prevent the carriers 
from spreading-out.

The purpose of this work is to show that the charge separation 
between the electron and the hole of a dot leads, just by itself, to 
a two-pair energy going above twice the one-pair energy. The 
analytical results presented here are very general, and apply to 
quantum dots of {\em any geometry} within the strong confinement 
regime : to use them for a particular experiment, one just has to 
introduce the specific carrier wave functions of the dot in the 
relevant quantities given in eqs(2,11,13). For the purpose of 
illustration, the numerical results given here correspond to a model 
spherical dot. In order to fully control the physics of this phenomenon, 
we have also reconsidered analytically the case of one pair plus one 
carrier~\cite{Lelong96}. Even without electric field, the energy of 
one pair plus one hole ends by going above the energy of one pair 
plus the energy of one hole if - but only if - the electron spreads 
out more than the hole, while in the case of two pairs, the electron 
and hole just have to spread out {\em differently}, for the 
''antibinding'' to appear.

\section{General background on a few carriers in quantum dot}

One carrier, electron {\em (e)}, or hole {\em (h)}, trapped in a dot, 
is characterized by a quantum number $n_i$, with $i=(e,h)$, its 
energy and wave function being $\enj{i}{n_i}$ and 
$\varphi_{n_i}^{(i)}(\rr)$. If we put more than one carrier in a dot, 
they feel each other by Coulomb interactions - and possibly by Pauli 
exclusion, if their spins are identical. The Coulomb potential in a 
confined geometry is characterized by a set of matrix elements 
$\vij{i}{j}{n'_i}{m'_j}{m_j}{n_i}$ between electrons, between holes 
and between electrons and holes, defined as
\begin{equation}\label{eq:Vij}
\vij{i}{j}{n'_i}{m'_j}{m_j}{n_i} =  \int d^3\rr \, d^3\rr' \,
{\varphi^{(i)}_{n'_i}}^*(\rr) \, {\varphi^{(j)}_{m'_j}}^*(\rr') \,
\frac{e^2}{\absv{\rr-\rr'}} \, {\varphi^{(j)}_{m_j}}(\rr') \,
{\varphi^{(i)}_{n_i}}(\rr)
\end{equation}
In small enough dots, it is well-known that the energy of a few 
carriers is dominated by the kinetic contribution, and so that the 
Coulomb interactions can be treated as a 
perturbation~\cite{Banyai88,Yoffe02}.  Up to second order, the ground 
state energy of one electron-hole pair thus reads as
\begin{equation}\label{eq:E0XExp}
{\cal E}^{(eh)}_{\bar{0}} = \ene{0}+\enh{0} - \veh{0}{0}{0}{0} +
W^{(e h)} + \cdots
\end{equation}
where $0$ is the ground state quantum number, the second order 
Coulomb term $W^{(i, j)}$ being
\begin{equation}\label{eq:secondO}
W^{(i j)} = \sum_{(n_i,m_j) \neq (0,0)}
\frac{\absv{\vij{i}{j}{n_i}{m_j}{0}{0}}^2}{\enj{i}{0}+\enj{j}{0}-\enj{
i}{n_i}-\enj{j}{m_j}}
\end{equation}
In the same way, the ground state energy of one pair plus one carrier 
$i=(e,h)$, with different spins, reads
\begin{equation}\label{eq:E0tExp}
E^{(e h i)}_{0} = \enj{e}{0} + \enj{h}{0} + \enj{i}{0} +
\vij{i}{i}{0}{0}{0}{0} - 2  \, \vij{e}{h}{0}{0}{0}{0} + W^{(i i)} + 2
\, W^{(e h)}
\end{equation}
while the ground state energy of two pairs with different spins is given by
\begin{equation}\label{eq:E0dbar}
E^{(eehh)}_{0} = 2  \, \ene{0} + 2  \, \enh{0} + \veeMA{0}{0}{0}{0} +
\vhh{0}{0}{0}{0} - 4  \, \veh{0}{0}{0}{0} + W^{(e e)} + W^{(h h)} + 4
\, W^{(e h)} + \cdots
\end{equation}

The Coulomb expansions of the carrier energies given above are valid 
when the dot size is small, more precisely when the dimensionless 
parameter $r_d$, characterizing a dot of volume $\Omega$, defined as
\begin{equation}\label{eq:rD}
\Omega = \frac{4}{3} \pi r_d^3 a_X^3
\end{equation}
is small compared to 1, $a_X=\hbar^2 / \mu e^2$ being the Bohr radius 
with $\mu^{-1} = m_e^{-1}+m_h^{-1}$. (For spherical dot , $r_d$ is 
just the dot radius in Bohr units). The Coulomb expansions 
(\ref{eq:E0XExp}-\ref{eq:E0dbar}), valid for small dots, in fact 
correspond to a small $r_d$ expansion.

Eqs.(\ref{eq:E0XExp},\ref{eq:E0dbar}) allow to obtain the lowest 
energies of one pair, two pairs and one pair plus one carrier for 
{\em any dot shape and barrier height}, up to second order in Coulomb 
interaction: to get them, we just need to first determine the free 
carrier eigenstates, $\enj{i}{n_i}$ and $\varphi^{(i)}_{n_i}(\rr)$ 
(see e.g.~\cite{Rodt03,Rodt05,Michelini04}), and then to use these 
wave functions in the $\vij{i}{j}{}{}{}{}$ Coulomb matrix elements 
defined in eq.(\ref{eq:Vij}).

For the purpose of illustration, we here consider a {\em model spherical dots 
with infinite barriers}. The problem is quite simple in the case of spherical 
dots because the free carrier eigenstates are then analytically known, the 
ground state energy being given by
\begin{equation}\label{eq:E0}
\enj{i}{0} = \frac{\pi^2}{r_d^2} \frac{\mu}{m_i} \, R_X
\end{equation}
with $R_X = \hbar^2/2 \, \mu \, a_X^2$. As the wave functions 
$\varphi_{n_i}^{(i)}(\rr)$ for infinite barriers do not depend on 
mass, the $\vij{i}{j}{n'_i}{m'_j}{m_j}{n_i}$'s do not depend on 
$(i,j)$, the one between ground states being equal to 
$\vij{i}{j}{0}{0}{0}{0} \simeq 3.57 \, R_X / r_d$. This makes all the 
second order Coulomb terms $W^{(i j)}$ also equal for equal electron 
and hole masses - while they differ for $m_e \neq m_h$. 

Consequently, in the case of spherical dots with infinite barriers, we find 
the following energy expansions:
\begin{eqnarray}
{\cal E}^{(eh)}_{0} &= & R_X \, \left[
\frac{\pi^2}{r_d^2}-\frac{3.57}{r_d} - c^{(eh)}(m_e,m_h) +  {\cal
O}(r_d) \right]   \nonumber  \\
E^{(e h i)}_{0} &= & R_X \, \left[ \frac{\pi^2}{r_d^2}
\left(1+\frac{\mu}{m_e} \right)-\frac{3.57}{r_d} - c^{(e h
i)}(m_e,m_h) + {\cal O}(r_d) \right]  \nonumber \\
E^{(eehh)}_0 &= & 2 \, R_X \, \left[
\frac{\pi^2}{r_d^2}-\frac{3.57}{r_d} - c^{(eehh)}(m_e,m_h) + {\cal
O}(r_d) \right]  \label{eq:E0dbarNum}
\end{eqnarray}
For $m_e = m_h$, all the W's are equal to $(-\gamma \, R_X)$ with 
$\gamma=0.133$ so that $c^{(eh)}=\gamma$, while $c^{(e h 
i)}=c^{(eehh)}=3 \, \gamma$ (Note that $E^{(eehh)}_0$ has a factor 2 
in front).
For different electron and hole masses, more precisely, in the 
particular case of $m_e=0.0665$ and $m_h=0.340$, which corresponds to 
pure $Ga As$, these quantities become $c^{(eh)}=0.182$, 
$c^{(ehh)}=0.772$, $c^{(ehe)}=0.444$ while $c^{(eehh)}=0.608$ (The 
first 20 electron and 20 hole levels were taken into account to 
achieve convergence of these sums).

\section{Carrier ''binding'' energy}

The ''binding'' energy $\Delta^{(e h i)}$ of one pair plus one 
carrier $i=(e,h)$ can be defined as
\begin{eqnarray}\label{eq:bindHtriX}
- \Delta^{(e h i)} &= & E^{(e h i)}_{0} - {\cal E}^{(e h)}_{0} -
\enj{i}{0} \nonumber  \\
&= & \delta_1^{(e h i)} + \delta_2^{(e h i)} + \cdots
\end{eqnarray}
Using eqs.(\ref{eq:E0XExp},\ref{eq:E0tExp}), we find that the second 
order term is just $\delta_2^{(e h i)} = W^{(e h)} + W^{(i i)}$ while 
the first order term can be rewritten~\cite{Lelong96}, using the 
definition of $\vij{i}{j}{0}{0}{0}{0}$ given in eq.(2), as
\begin{equation}\label{eq:d1iij}
\delta_1^{(e h i)} = \int d\rr \, d\rr' \,
\frac{e^2}{\absv{\rr-\rr'}} \, n_i(\rr) \,
\absv{\varphi^{(i)}_0(\rr')}^2
\end{equation}
where $n_i(\rr) = n(\rr) = \absv{\varphi^{(h)}_0(\rr)}^2 - 
\absv{\varphi^{(e)}_0(\rr)}^2 $ for $i=h$ and $n_i(\rr) = -n(\rr)$ 
for $i=e$.

In the same way, the ''binding'' energy of two pairs can be defined as
\begin{eqnarray}\label{eq:bindHbiX}
- \Delta^{(eehh)} &= & E^{(eehh)}_{0} - 2 \, {\cal E}^{(eh)}_{0}  \nonumber  \\
&= & \delta_1^{(eehh)} + \delta_2^{(eehh)} + \cdots
\end{eqnarray}
When using eqs.(\ref{eq:E0XExp},\ref{eq:E0dbar}), the second order 
term is just $\delta_2^{(eehh)} = W^{(e e)}+W^{(h h)}+2 \, W^{(e h)}$ 
while the first order term now reads
\begin{equation}\label{eq:d1eehh}
\delta_1^{(eehh)} = \int d\rr \, d\rr' \, \frac{e^2}{\absv{\rr-\rr'}}
\, n(\rr) \, n(\rr')
\end{equation}

From Eqs.(\ref{eq:d1iij},\ref{eq:d1eehh}) and the definitions of the 
$\delta$'s, it is easy to check that a remarkable sum rule exists between 
the "binding energies" of two pairs and of one pair plus one carrier:
\begin{eqnarray}
\delta_1^{(eehh)} &= & \delta_1^{(ehe)} + \delta_1^{(ehh)}  \nonumber \\
\delta_2^{(eehh)} &= & \delta_2^{(ehe)} + \delta_2^{(ehh)} \label{eq:sumBE}
\end{eqnarray}

Let us stress that Eqs.(\ref{eq:d1iij},\ref{eq:d1eehh}) as well 
as~Eq.(\ref{eq:sumBE}) are completely general, i.e., they {\em do not rely on 
any specific assumption for the dot geometry nor on a possibly non-zero 
electric field}. From Eqs.(\ref{eq:d1iij},\ref{eq:d1eehh}) we already see that 
the first order Coulomb terms of these ''binding'' energies reduce to zero 
if $n(\rr) = 0$ everywhere, i.e., if the dot has a local carrier neutrality. 

\section{Dot with local carrier neutrality}

Local carrier neutrality implies the absence of any external electric 
field which tends to tear apart opposite charges. We also need to 
assume infinite barriers or, possibly, carriers spreading out of the 
dot identically, for their wave functions to be the same.

For $n(\rr) = 0$, the first order terms, $\delta_1^{(e h i)}$ and 
$\delta_1^{(eehh)}$ reduce to zero~\cite{Lelong98}. If we now turn to the second 
order terms, $\delta_2^{(e h i)}$ and $\delta_2^{(eehh)}$ , we see 
that they are both negative, for all the W's are negative, the sum 
they contain being taken over excited states. These second order 
terms, which are the dominant ones in small dots in the absence of 
first order terms, make the two binding energies $\Delta^{(e h i)}$ 
and $\Delta^{(eehh)}$ positive (for the latter case, see~\cite{Banyai88}). 
We conclude that, in a small dot with infinite barrier, two-pairs, 
and one-pair plus one carrier, are always below the ''dissociated'' 
configuration, i.e., twice one-pair or one-pair and one carrier.

\section{Dot with local charge separation}

For non-zero electric fields, or for finite barriers and different 
masses, i.e., different $(m_i, V_i)$, the two types of carriers 
generally have different wave functions, so that $n(\rr)$ differs 
from zero. Due to $e^2 / \absv{\rr-\rr'}$, the integrals of 
$\delta_1^{(e h i)}$ and $\delta_1^{(eehh)}$, in 
eqs.(\ref{eq:d1iij},\ref{eq:d1eehh}), are dominated by the $\rr 
\simeq \rr'$ domain. As for such $(\rr, \rr')$, we have $n(\rr) \, 
n(\rr') \simeq [n(\rr)]^2$, so that the integrand of 
$\delta_1^{(eehh)}$ is positive in the relevant part of the integral, 
whatever the sign of $n(\rr)$, making $\delta_1^{(eehh)}$ always 
positive.

If we turn to $\delta_1^{(e h i)}$, we see that, due to the 
additional $\absv{\varphi^{(i)}_0(\rr)}^2$, the important part of the 
integral given in eq.(\ref{eq:d1iij}), is now the one for $\rr \leq 
R$. Consequently, the sign of $\delta_1^{(e h i)}$ is controlled by 
the sign of $n_i(\rr)$ inside the dot. As the electron is usually the 
carrier which spreads out the more, the hole wave function in the dot 
is larger than the electron one, for the wave functions are 
normalized. This leads to $n(\rr)$ essentially positive in the dot, 
making $\delta_1^{(ehh)}$ positive and $\delta_1^{(ehe)}$ negative.

When the first and second order terms are both negative, as for 
$(ehe)$, the carrier ''binding'' energy is unambiguously positive, 
even for extremely small dots. On the opposite, when the first order 
term is positive, as for $(eehh)$ and $(ehh)$, this first order term 
- even if it is very small, i.e., if the electron and hole nearly 
have the same wave function - must end by being the dominant Coulomb 
contribution when the dot shrinks. Consequently, the carrier 
''binding'' energy, positive for intermediate dot sizes - as it is 
then dominated by the second order Coulomb term - must turn negative 
when the dot shrinks, in qualitative agreement with experimental 
data~\cite{Rodt03,Rodt05}. Therefore the phenomenon of competition 
between first and second order Coulomb contributions drives the cross-over 
between binding and antibinding. In~\cite{Lelong98} we find a numerical 
calculation up to second order in the Coulomb interaction illustrating 
ideally our argument. One even notices that our 
sum rule~(\ref{eq:sumBE}) is accurately verified by Fig.2 of~\cite{Lelong98} 
in most of the size range (namely above $r=90A$). Unfortunately in 
the antibinding region, below $r=90A$, a small discrepancy appears, 
probably due to limitations in the calculation of the second order term. 
Nevertheless the overall numerical result of Fig.2 beautifully confirms 
the findings of our analytical theory.

To conclude we state our main thesis which says that, in order to 
find an ''antibinding'' for 
two-electron-hole pairs, we just need $n(\rr) \neq 0$, i.e., a 
carrier local non-neutrality, while to find such an ''antibinding'' 
for one-pair plus one carrier, we need an excess charge inside the 
dot of the same sign than the additionnal carrier. This conclusion 
fully agree with experimental data~\cite{Regelman02,Besombes02,Ashmore02}.

\section{Link with the carrier spreading-out}
Let us end this work by taking again for an illustration, a quantum 
dot with a spherical geometry, and show how we can relate the dot 
size for the cross-over from ''binding'' to ''antibinding'' of 
$(eehh)$ and $(ehh)$, to one of the important physical quantities for 
carriers in dots, namely their spreading-out lengths.

In a previous communication~\cite{Leyronas01}, we have shown that the 
energies of a particle with mass $m_i$ in a spherical dot of radius 
$R$ and barrier height $V_i$, are given by $\alpha_i^2\hbar^2 / 2 m_i 
R^2\equiv\alpha_i^2 \, R_X (\pi^2 / r_d^2) (\mu / m_i )$. The 
$\alpha_i$'s for states with $l=0$ symmetry fulfil $\nu_i = \alpha_i 
/ \sin(\alpha_i)$, where $\nu_i$ is the parameter defined 
in Eq.(\ref{eq:Vi}). In the large $\nu_i$ limit, i.e., for large 
$V_i$, this leads to $\alpha_i \sim \pi / (1+\nu_i^{-1})$ for the 
ground state; so that the spatial extension $d_i$ of this ground 
state, defined as $E_i =\hbar^2 / 2 m_i d_i^2$, varies with the 
effective barrier height $\nu_i$ as $d_i \simeq R \, (1+\nu_i^{-1})$. 
Note that, as expected, $d_i$ is just equal to $R$ for infinite 
barriers, i.e., for infinite $\nu_i$.

We now use this result in the ''binding'' energy first order terms, 
Eqs.(\ref{eq:d1iij},\ref{eq:d1eehh}): since, due to dimensional arguments, 
$\absv{\varphi^{(i)}}^2 \simeq 1 / d_i^3$, the first order term 
$\delta_1^{(eehh)}$, given in eq.(\ref{eq:d1eehh}), can be estimated 
as
\begin{eqnarray}
\delta_1^{(eehh)} & \simeq & R^3 R^3 \frac{e^2}{R} \left(
\frac{1}{d_h^3} - \frac{1}{d_e^3} \right)^2 \nonumber \\
& \simeq & \frac{e^2 \, (d_e-d_h)^2}{R^3} \simeq \frac {e^2 \,
(\nu_e^{-1}-\nu_h^{-1})^2}{R^3} \label{eq:estd1eehh}
\end{eqnarray}
while the same argument leads to
\begin{equation}
\delta_1^{(e h h)} \simeq \frac{e^2 \, (\nu_e^{-1}-\nu_h^{-1})}{R}
\label{eq:estd1iij}
\end{equation}
with a similar result for $\delta_1^{(e h e)}$.

We now define the characteristic length $l_i$ over which a carrier 
$m_i$ spreads out of a material having a barrier $V_i$, as
\begin{equation}
V_i = \frac{\hbar^2}{2 m_i l_i^2}
\end{equation}
(Note that this $l_i$ is inversely proportional to $\sqrt{m_i V_i}$, 
while it is exactly $0$ for infinite barrier). Following part I, the 
second order Coulomb term is of the order of $(-e^2 / a_X)$, so that, 
from the definition of $\nu_i$ given in eq.(\ref{eq:Vi}) - in which 
enters the dot radius - we obtain a cross-over radius from 
''binding'' to ''antibinding'' which behaves as
\begin{eqnarray}
R^{(eehh)} & \simeq & \sqrt[3]{a_X (l_e-l_h)^2} \label{eq:Rcriteehh} \\
R^{(e h i)}& \simeq & \sum_{j=(e,h)} \Theta(l_j-l_i) \sqrt{a_X
(l_j-l_i)} \label{eq:Rcritiij}
\end{eqnarray}
where $\Theta(x)$ is the step function. This gives a finite 
cross-over radius for $(eehh)$ whatever ($l_e, l_h$) are, while the 
one for $(e h i)$ depends on the sign of $(l_e-l_h)$. For 
$l_e-l_h>0$, which is the most usual situation, the cross-over radius 
for $(ehh)$ is finite while the one for $(ehe)$ is zero, i.e., no 
cross-over takes place when the dot is negatively charged.

Eqs.(\ref{eq:Rcriteehh}-\ref{eq:Rcritiij}) also show that when the 
barriers are very high, the  spreading-out lengths $l_i$ are very 
small, so that the cross-over radii are very small. For usual barrier 
heights, however, the $l_i$'s are of the order of the Bohr radius 
$a_X$, making the cross-over radii also of the order of $a_X$. In 
order to fit a particular experiment, it is possible to get precise 
values of these cross-overs by going back to the expressions of the 
energies given in eqs.(\ref{eq:E0XExp}-\ref{eq:E0dbar}), the purpose 
of this last part being just to get {\em a physical understanding of 
this cross-over by establishing its physical link with the carrier 
spreading-out lengths}.

One should not however conclude in all cases that charge separation 
increases when the quantum dot size diminishes. For example in 
wurtzite-type GaN/AlGaN heterostructures, where piezoelectricity 
or spontaneous polarization are prominent effects, charge separation 
effects may increase with the quantum dot size~\cite{Grund95}, therefore 
the behaviour of ''binding energies'' with the box size may be 
strongly affected.

\section{Comparison with other approaches}

A number of authors have made very complex calculations of 3D wave 
functions (accounting for the details of the confinement potential 
resulting from the inhomogeneous strain, band mixing, and the 
piezolelectric potential), and subsequently have carried out 
configuration-interaction calculation of the biexciton binding 
energy. Although it is not our purpose here to include such effects, our 
approach is able fully exploit the results of any such complex 3D 
numerical single particle wave functions: the contrast lies in the analysis 
of the results. An evaluation of Eq.(\ref{eq:bindHbiX}) with such wave 
functions allows to firmly assess the exact size limit for the 
validity of the strong confinement regime: for that, we just have to 
compare the level shifts of the two approaches. More important, 
Eq.(\ref{eq:bindHbiX}) also allows to assess the relative magnitude of the 
first and second order Coulomb contributions for different dot sizes. Note that
in the second order contribution, can also enter a nearby continuum 
of states. Finally, a numerical evaluation of Eq.(\ref{eq:d1eehh}) allows {\em 
to prove} that charge separation is already of importance at first order, and 
being actually the {\em main cause} for the antibinding of two pairs.

Let us now show how the results presented here, which are completely 
general, would actually bring useful insights in the understanding of 
specific experiments.

We focus on Refs.~\cite{Rodt03,Rodt05} where the 
transition from binding and antibinding is systematically studied, 
both experimentally and numerically. These authors find a qualitative 
agreement with experiments when the aspect ratio is varied, but not 
the dot size. Their results also show that, in the two-pair ground 
state of the largest dot, namely 20 nm, there is still a relatively 
small mixing with the other excited states due to the Coulomb 
interaction, showing in this way that, in smaller dots, the strong 
confinement regime is certainly reached. The antibinding is then 
attributed to a number of combined effects such as ''3D confinement, 
quenching correlations and exchange, and causing local charge 
separation'', without precise evaluation of their relative importance, 
this relative importance being however crucial for physical understanding.

In order to show how we can analyze the results of the numerical 
approaches within our procedure, let us focus on the calculation 
presented in~\cite{Rodt03}. In this work, the authors do not vary the 
dot size to understand the transition to antibinding - which is the 
physically relevant parameter - but vary the number of confined 
states they include in the sums - which only is a mathematically 
relevant parameter. Indeed, their numerical procedure is (i) {\em to 
fix the dot size at 13 nm} and (ii) {\em to vary the number of bound 
states taken into account} in the calculation, between 1 and 3. From 
our approach, it is clear that there are fundamental flaws in this 
procedure: indeed, the confinement energy and the first and second 
order Coulomb contributions all have {\em a different}, but crucial, 
{\em dot size dependence} (see e.g.~the explicit $r_d$ dependence in 
Eq.(\ref{eq:E0dbar})) these dependences having nothing to do with 
the possible variation of the number of confined states included in 
the numerical calculation. The latter procedure amounts to only 
change the magnitude of the second order terms, {\em without any size 
effect}. 

A refined set of calculations is presented in Ref.~\cite{Rodt05} where the previous 
criticism do not fully apply. Indeed the authors convincingly show that a complex CI 
calculation reproduces the trend of the experiments when one truly varies the 
QD size. They attribute the crossing to ''correlation'' (which we here simply 
call ''second order corrections''). The authors of Ref.~\cite{Rodt05} 
check that the number of excited hole bound states affect the crossing, 
whilst the electrons do not. We agree and think that this 
is a natural result of the smaller hole level spacing. 
However besides the convergence check we feel again that it is difficult to 
draw definitive {\em physical} conclusions about the actual number of bound 
states from this artificial procedure. In particular we note that there 
are also in principle contributions from the continuum, and when a higher 
bound state ''disappears'', it merges in the continuum, however it is not 
obvious to guess how much this effect increases the contribution of the 
continuum. 

The ''Quantum Confined Stark Effect'' on one and two-pair states in 
small dots has also been investigated in two different sets of 
experiments namely, random local field~\cite{Besombes02}, or external 
field~\cite{Sugisaki02}. Both show that the binding energy of 
two-pairs decreases with the external electric field strength at the 
dot position. Such a behaviour is in perfect agreement with our 
discussion concerning the importance, the sign and magnitude of the 
first order term~(\ref{eq:d1eehh}) in small dots, as a function of 
charge separation.

Experiments~\cite{Regelman02,Besombes02,Ashmore02}, involving the 
states of one pair plus one carrier (the so-called ''charged 
excitons''), with possibly an additional external field 
~\cite{Besombes02,Ashmore02}, show that, in small dots, the binding 
energy is of opposite sign for the two types of excess charge, and 
that the trend to ''antibind'' is enhanced by the field for both 
types of excess charge. The authors explain it qualitatively by 
saying that the electric field tends to tear apart opposite charges 
and keep together identical charges, so that the repulsive Coulomb 
interactions are wining over the attractive ones when the field 
increases. This first explanation is fully intuitive. Our 
Eq.~(\ref{eq:d1iij}) shines new light on this problem because it 
demonstrates that, in the end, it is just this exact integral involving 
only the charge separation, evaluated with single-particle 
wavefunctions, that matters to understand the behaviour of the 
''binding energy''.

\section{Conclusion}

We have shown, in very general terms, that the two-pair ground state 
energy, in strongly confined quantum dots, can possibly go above 
twice the energy of one-pair due to a single physical quantity: the local 
charge separation. Our conclusion holds independently of the physical origin 
of the charge separation, which can be complex and internal (e.g. due to 
piezoelectric fields resulting from strain), or external (e.g. applied 
electric fields). Even in the absence 
of electric field, local charge separation can be induced by finite barrier 
heights, the carriers spreading out of the dot differently. Only the precise 
value of the cross-over is influenced by the complicated geometry of real 
dots. It is attributed to a competition effect between the first and 
second order Coulomb contributions. While such an ''antibinding'' 
always exists for two-pairs, it only exists for one-pair plus one 
carrier if the additionnal carrier is the one which spreads out the 
less. For illustration, we have, in the case of spherical dots, 
related the radius of the cross-over  from ''binding'' to 
''antibinding'' to the typical carrier spreading-out lengths induced 
by the finite dot barriers. As a by-product we have also found a remarkable 
sum rule for the ''binding energies'' of two pairs and one pair plus one carrier. 
Finally, we have shown how our approach 
can be used to analyse the results of complex numerical calculations 
of two-pair states in realistic dot geometry, and how it allows to 
reinterpret a variety of experimental data in strongly confined 
quantum dots.

\end{document}